\documentclass[journal,comsoc]{IEEEtran}

\usepackage{amsmath}
\usepackage{amsthm}
\usepackage{amsfonts}
\usepackage{stackrel}
\usepackage{color}
\usepackage{cite}
\usepackage{hyperref}
\usepackage[T1]{fontenc} % optional
\usepackage[cmintegrals]{newtxmath}
\usepackage{bm} % optional

\newcommand{\ud}{\,\mathrm{d}}
\newcommand{\ue}{\mathrm{e}}
\newcommand{\bx}{\mathbf{x}}
\newcommand{\by}{\mathbf{y}}
\newcommand{\bX}{\mathbf{X}}
\newcommand{\bY}{\mathbf{Y}}
\newcommand{\bZ}{\mathbf{Z}}
\newcommand{\bR}{\mathbb{R}}
\renewcommand{\Pr}[1]{\mathrm{Pr}\left\lbrace{#1}\right\rbrace}
\newcommand{\h}[1]{\mathrm{h}\left( {#1} \right)}
\newcommand{\E}[1]{\mathbb{E}\left[{#1}\right]}

\newtheorem{definition}{Definition}
\newtheorem{theorem}{Theorem}
\newtheorem{lemma}{Lemma}
\newtheorem{remark}{Remark}
\newtheorem{example}{Example}

\begin{document}
%%%%%%
% paper title
\title{Existence and Continuity of Differential Entropy for a Class of Distributions}

%%%%%
% Authors
\author{
	Hamid~Ghourchian,
	Amin~Gohari,~\IEEEmembership{Senior~Member,~IEEE}, and
	Arash~Amini,~\IEEEmembership{Senior~Member,~IEEE}

	\thanks{}

	\thanks{Manuscript received November 17, 2016; revised February 22, 2017; accepted March 22, 2017. The work of A. Gohari was supported by the Sharif University of Technology under Grant QB950607.}

	\thanks{The authors are with the Department of Electrical Engineering and Advanced Communication Research Institute (ACRI), Sharif University of Technology, Tehran, Iran. E-mails: h\_ghourchian@ee.sharif.edu, \{aminzadeh, aamini\}@sharif.ir.}
}

\maketitle

%%%%%
% Abstract
%%%%%
\begin{abstract}
%\boldmath
In this paper, we identify a class of absolutely continuous probability distributions, and show that the differential entropy is uniformly convergent over this space under the metric of total variation distance.
One of the advantages of this class is that the requirements could be readily verified for a given distribution.
\end{abstract}


\IEEEpeerreviewmaketitle

%%%%%%%
% Introduction
%%%%%%%
\section{Introduction}
\IEEEPARstart{T}{he} differential entropy of a given probability distribution represented by the density $p(x)$ is defined as
$
	\h{p}:=-\int_\bR{p(x)\log p(x)\ud x},
$
if the integral converges (\emph{existence of entropy}).
First defined by Shannon in \cite{Shannon48},  the differential entropy of a random variable is derived by subtracting $\log m$ from the discrete entropy of the quantized version of the random variable with step size $1/m$, and taking its limit as $m$ tends to infinity \cite[p. 247]{Cover06}.

Assuming the existence of entropy for a convergent sequence of probability measures (in some metric), it is of importance to determine whether the differential entropies also converge (\emph{continuity of entropy}).
Both existence and continuity of entropy has received attention in the literature \cite{Barron92, Poly16} due to their numerous applications, \emph{e.g.} see \cite{Barron92}.
To guarantee these two properties, it is common to impose constraints on the probability density function (pdf), or alternatively, to convolve the given distribution with a normal distribution to smooth it out.
For the latter approach see for instance the proof of the central limit theorem in \cite{Barron84}, and check \cite{Poly16} for an application of this approach in network information theory. 

In order to sketch the current landscape, we provide a non-rigorous statement of the existing results, without giving the exact technical conditions.
Assume a sequence of absolutely continuous ($\mathcal{AC}$) probability measures (w.r.t. Lebesgue measure) with pdfs $p_k(x)$.
We say that $p_k(x)$ converges to the pdf $p(x)$ in relative entropy if $D(p_k\Vert p) \to 0$, where $D(\cdot\|\cdot)$ is the relative entropy between the two pdfs, defined as 
$
	D(p\|q)=\int_{\mathbb{R}}{p(x)\log[{p(x)}/{q(x)}]\ud x}.
$
We say that the convergence holds in total variation if the total variation distance $\lVert p_k-p\rVert_1$ vanishes as $k\to \infty$, where
\begin{equation}
	\left\lVert p-q \right\rVert_1
	=\int_{\mathbb{R}}{\left|p(x)-q(x)\right|\ud x}.
\end{equation}
The convergence holds in entropy if ${\h{p_k}}\to\h{p}$, and finally, the convergence holds in ratio if ${p_k(x)}/{p(x)}\to 1$ for all $x$ in the support of $p$.
Pinsker's inequality shows that the convergence in relative entropy implies convergence in total variation \cite[p. 33]{Ihara93}. 
Provided $m<p_k(x), \,p(x)<M$ for all $k$ and $x\in\text{supp}\{p\}$, and some positive $m,\,M$, it is shown in \cite{Piera09} that the convergence in relative entropy implies the convergence in entropy.
Similarly, with $m<{p_k(x)}/{p(x)}<M$ one can show that the convergence in total variation implies the convergence in entropy. In \cite{Godavarti04}, under similar conditions, it is shown that the pointwise convergence of distributions implies their convergence in entropy. 
In \cite{Sason15, Piera09} it is shown that the convergence in ratio accompanied with $p_k(x)/p(x)<M$ for some $M>0$ implies the convergence in entropy and relative entropy.
A technique devised in \cite[p. 12]{Johnson04} reveals that if $p$ is the ``projection'' of some distribution $q$ onto a closed and convex set $\mathcal{F}$ of distributions, then, the convergence of a sequence in $\mathcal{F}$ in either entropy or total variation implies the convergence in relative entropy.
With uniform point-wise convergence of both $p_k\to p$ and $p_k/p\to 1$, the convergence in entropy is obtained in \cite{Ikeda59}.
Further, the condition on $p_k/p$ could be dropped if the distributions are of finite support.
It is shown in \cite{Otahal94} that the convergence in entropy holds if $p_k(x)^{1-\delta}$, $p_k(x)^{1+\delta}$, $p(x)^{1-\delta}$, and $p(x)^{1+\delta}$  are all integrable over the real line for some $0<\delta<1$, and the integrals of $|p_k(x)-p(x)|^{1+\delta}$ and $|p_k(x)-p(x)|^{1-\delta}$ vanish as $k\rightarrow \infty$.

Finally, a class of ``regular probability distributions" for \emph{vectors} of random variables is identified in \cite{Poly16}, for which the entropy difference of any two members could be controlled by their Wasserstein distance and variance.

In this paper we define the new class $(\alpha,v,m)\text{--}\mathcal{AC}^{n}$ of $n$-dimensional probability measures for which the convergence in total variation leads to the convergence in entropy.
The parameters $v$ and $m$ in this class represent upper bounds on the $\alpha$th order moment and the supremum of $p(x)$, respectively, which are easier to evaluate than the integrals of $p(x)^{1-\delta}$ and $p(x)^{1+\delta}$. 
This class is a subset of the one considered in \cite{Nair06}, for which the existence (and not continuity) of the differential entropy is shown.
Thus, with more stringent conditions, we prove continuity of entropy in addition to its existence.
Conversely, the authors of \cite{Poly16} define a class of ``$(c_1,c_2)$-regular distributions'' as the set of  $p(\bx)$ satisfying 
\begin{equation} \label{eqn:regular}
	\| \nabla \log p(\bx) \|_2 \leq c_1 \| \bx \|_2+c_2,
	\qquad \forall \bx\in\bR^n.
\end{equation}
The above regularity condition is restrictive. For instance, any regular $p(\bx)$ needs to be differentiable, whereas $p(\bx)\in (\alpha,v,m)\text{--}\mathcal{AC}^{n}$ may not even be continuous.
However, with the restrictive regularity condition, the scaling behaviour of the results in \cite{Poly16} in terms of $n$ are  superior than those of $(\alpha,v,m)\text{--}\mathcal{AC}^{n}$: it is proved in \cite[Proposition 1]{Poly16}  that if $\bX\sim p_\bX(\bx)$ and $\bY\sim p_\bY(\by)$ are $(c_1 ,c_2)$-regular, then $\h{\bX}$ and $\h{\bY}$ exist and 
\begin{equation} \label{eqn:Wasser hX-hY}
	|\h{\bX}-\h{\bY}| \leq \Delta,
\end{equation}
where $\Delta$ is a constant that grows like $\sqrt{n}$ as we increase the dimension $n$.
However, the constant that we find for $(\alpha,v,m)\text{--}\mathcal{AC}^{n}$ grows like $n\log n$.

This paper is organized as follows: in Section \ref{sec.Prelim}, some definitions are given; the main result is presented in Section \ref{sec.MainRes}, and its proof is given in Section \ref{sec:Proofs} and the appendix.

%%%%%%%%%%%%
% Definitions and Notation
%%%%%%%%%%%%
\section{Definitions and Notations} \label{sec.Prelim}
All the logarithms in this paper are in base $\ue$.
Random variables and vectors are denoted by capital letters with vectors distinguished by bold letters.

% Definition: Absolutely Continuous Random Vector
\begin{definition}[Absolutely Continuous Random Vector] \label{def.ACRV} 
	Let $\mathcal{B}^n$ be the Borel $\sigma$-field of $\mathbb{R}^n$ and let $\bX$ be a real-valued and $n$-dimensional random vector that is measurable with respect to $\mathcal{B}^n$. 
	We call $\bX$ an \emph{absolutely continuous} random vector if its probability measure $\mu$, induced on $(\mathbb{R}^n,\mathcal{B}^n)$, is absolutely continuous with respect to the Lebesgue measure for $\mathcal{B}^n$ (i.e., $\mu(\mathcal{A})=0$ for all $\mathcal{A}\in\mathcal{B}^n$ with zero Lebesgue measure). 
	We denote the set of all absolutely continuous distributions by $\mathcal{AC}$. 
\end{definition}
	
The Radon-Nikodym theorem implies that for each $\bX\in \mathcal{AC}$ there exists a $\mathcal{B}^n$-measurable function $p:\mathbb{R}^n\to [0,\infty)$, such that for all $\mathcal{A}\in \mathcal{B}^n$ we have that
\begin{equation}
	\Pr{\bX\in \mathcal{A}} = \int_\mathcal{A}{p(\bx) \ud \bx}.
\end{equation}
The function $p$ is called the probability density function (pdf) of $\bX$ \cite[p. 21]{Ihara93}.
The property $\bX\in\mathcal{AC}$ is alternatively written as  $p\in\mathcal{AC}$. Random variables with an absolutely continuous distributions are generally known as  continuous random variables.

% Remark: Absolute Continuity of functions
\begin{remark}
	Intuitively speaking, the absolute continuity of a probability measure \emph{w.r.t} the Lebesgue measure implies that the function has a pdf with no delta functionals, \emph{i.e.,} the cumulative distribution function (CDF) is continuous.
	It does not imply the absolute continuity (or even continuity) of its probability density function $p(\bx)$ as a function of $\bx$.
\end{remark}

% Definition: Differential Entropy
\begin{definition}[Differential Entropy] \label{def.Ent} \cite[Chapter 2]{Cover06}
	For an $n$-dimensional $\bX\in\mathcal{AC}$ with density $p$
	we define the differential entropy $\h{\bX}$, or $\h{p}$ as
	\begin{equation}
		\h{p}=\h{\bX} := -\int_{\mathbb{R}^n}{p(\bx)\log {p(\bx)}\ud \bx},
	\end{equation}
	if the integral converges. We use the convention $x\log(x)=0$ at $x=0$.
\end{definition}

% Definition: (a,m,v)-AC
\begin{definition} \label{def.(a,m,v)-AC}
Given $\alpha,m,v>0$, and $n\in\mathbb{N}$, we define $(\alpha,v,m)\text{--}\mathcal{AC}^{n}$ to be the class of all $\bX\in\mathcal{AC}$ such that the corresponding density function $p:\mathbb{R}^n\mapsto[0,\infty)$ satisfies
	\begin{align}
		&\int_{\mathbb{R}^n}{\lVert \bx \rVert_\alpha^\alpha \, p(\bx)\ud \bx}<v, ~ {\rm and} \label{eqn2:defAC} \\
		& {\rm ess}\sup_{\bx\in\bR^n} ~p(\bx) < m, \label{eqn1:defAC}
	\end{align}
	where $\lVert \bx \rVert_\alpha := \big(\sum_{i=1}^n{\left|x_i\right|^\alpha}\big)^{1/\alpha}$.
\end{definition}

% Remark
\begin{remark}
	Observe that a scalar continuous random variable $X$ is in $(\alpha,m,v)-\mathcal{AC}$ if it has an $\alpha$-moment $\mathbb{E}[|X|^\alpha]<v$  and a bounded density function $p_X(x)<m$.
It is often not difficult to verify these properties for a given random variable.
	Many commonly used random variables have bounded density functions and at least one finite moment.
	Furthermore, if $X_1\in(\alpha,m_1,v_1)-\mathcal{AC}$ and $X_2\in(\alpha,m_2,v_2)-\mathcal{AC}$ are independent random variables, $X_1+X_2\in (\alpha,\min(m_1, m_2),c_{\alpha}(v_1+v_2))-\mathcal{AC}$ where $c_{\alpha}=1$ if $\alpha\in[0,1]$ and $c_{\alpha}=2^{\alpha-1}$ for $\alpha>1$.
	This is because $f_{X_1+X_2}(x)=\int_{t}f_{X_1}(t)f_{X_2}(x-t)dt\leq m_1\int_{t}f_{X_2}(x-t)dt=m_1$.
	Also, the $\alpha$-moment of $X_1+X_2$ can be bounded by the moments of $X_1$ and $X_2$ via the $c_r$-inequality for moments.
	Thus, any linear combination of the commonly used variables also belongs to the defined class of distributions.
\end{remark}

% Definition: generalized normal distribution
\begin{definition} \cite{Goodman73} \label{def.PhiAlphaV}
	The density of an $n$-dimensional random vector of \lq\lq$\alpha$-generalized normal distribution\rq\rq, $\mathcal{N}\left(0,(v\alpha/n)^{1/\alpha}\mathbf{I}_n,\alpha\right)$, is given by
	\begin{equation}
		\phi_{n,\alpha,v}(\bx)
		=\left[\left(\tfrac{n}{\alpha v}\right)^\frac{1}{\alpha}\tfrac{1}{2\Gamma\left(\frac{1}{\alpha}+1\right)}\right]^n
		\ue^{-\frac{n}{\alpha v} \lVert\bx\rVert_\alpha^\alpha},
	\end{equation}
	where $\mathbf{I}_n$ is the identity matrix of size $n$, scalars $\alpha,v>0$ are arbitrary, and $\Gamma$ is the {gamma function}.
	The $\alpha$-generalized normal distribution satisfies $\E{\left\lVert\bX_{n,\alpha,v}\right\rVert_\alpha^\alpha}=v$.
\end{definition}

%%%%%%%
% Main Results
%%%%%%%
\section{Main Result} \label{sec.MainRes}
Our main result stated below in Theorem \ref{thm.EntConv} implies that the differential entropy is uniformly continuous over $(\alpha,v,m)\text{--}\mathcal{AC}^{n}$ with respect to the total variation metric:
\begin{equation}
	\left\lVert p_\bX - p_\bY \right\rVert_1
	=\int_{\mathbb{R}^n}{\left|p_\bX(\bx)-p_\bY(\bx)\right|\ud \bx}.
\end{equation}

% Theorem: Entropy Convergent for Finite Variances
\begin{theorem} \label{thm.EntConv}
	The differential entropy of any $p_\bX\in(\alpha,v,m)\text{--}\mathcal{AC}^{n}$ is well-defined and satisfies
	\begin{align}
		|\h{p_\bX}| \leq&
		\tfrac{n}{\alpha}\left|\log\tfrac{\alpha v}{n}\right|
		+\log(\max\lbrace 1,m \rbrace) \nonumber\\
		&+n\log\left[2\Gamma\left(\tfrac{1}{\alpha}+1\right)\right]
		+\tfrac{n}{\alpha}+1.\label{eq.dh(pd)=0}
	\end{align}
	Furthermore, for all $p_\bX,p_\bY\in(\alpha,v,m)\text{--}\mathcal{AC}^{n}$ {satisfying $\left\lVert p_\bX - p_\bY \right\rVert_1\leq m$, we have that}
	\begin{align}
		&\left|\h{p_\bX}-\h{p_\bY}\right| \leq\nonumber\\
		&\quad c_1 \left\lVert p_\bX - p_\bY \right\rVert_1 + c_2 \left\lVert p_\bX - p_\bY \right\rVert_1 \log\frac{1}{\left\lVert p_\bX - p_\bY \right\rVert_1} \label{eqn:h-h<l1},
	\end{align}
	where
	\begin{align}
		c_1 =&\tfrac{n}{\alpha}\left|\log\tfrac{2\alpha v}{n}\right| + |\log(m \ue)|
		+\log\tfrac{\ue}{2}\nonumber \\
		&+n\log\left[2\Gamma\left(1+\tfrac{1}{\alpha}\right)\right]\label{eqc1}
		+ \tfrac{n}{\alpha}+1, \\
		c_2 =& \tfrac{n}{\alpha} + 2.\label{eqc2}
	\end{align}
\end{theorem}

We now compliment the above result by showing that the differential entropy may no longer be well-defined if we remove any of  \eqref{eqn2:defAC} or \eqref{eqn1:defAC} from the definition of $(\alpha,v,m)\text{--}\mathcal{AC}^n$.

% Example: h=+inf
\begin{example}
	Consider the following \iffalse non-negative-valued function \else pdf \fi defined over the real line $\bR$:
	% +m AC -a v
	\begin{equation}
		p(x)=
		\begin{cases}
			\frac{1}{x(\log x)^2}, & x>\ue, \\
			0, & x \leq \ue.
		\end{cases}\label{def-p-eq}
	\end{equation}
Note that $p(x)$ is a valid density function with no delta functionals; it defines a continuous random variable and thus $p\in\mathcal{AC}$. Also, observe that $p(x)<1/\ue$ for all $x\in\mathbb{R}$.
	On the other hand, by applying the change of variables $y=\log x$, we observe that the $\alpha$th-order moment is infinite for any $\alpha>0$:
	\begin{equation}
		\int_\ue^\infty{\frac{x^\alpha}{x (\log x)^2}\ud x}
		= \int_1^\infty{\frac{\ue^{\alpha y}}{y^2}\ud y}
		= \infty.
	\end{equation}
	Hence, this distribution satisfies \eqref{eqn1:defAC}, but not \eqref{eqn2:defAC}.
	With the same change of variables $y=\log x$, it can be observed that
	\iffalse
	\begin{align}
		\h{p}=&\int_\ue^\infty{\frac{1}{x(\log x)^2}\log(x(\log x)^2) \ud x} \\
		=&\int_1^\infty{\frac{1}{y^2}\log(y^2 \ue^y) \ud y}\\
		=&2\int_1^\infty{\frac{\log y}{y^2} \ud y}
		+\int_1^\infty{\frac{1}{y} \ud y}=+\infty.
	\end{align}
	where $y=\log x$.
	\else
	\begin{equation}
		\h{p}=\int_\ue^\infty{\frac{1}{x(\log x)^2}\log(x(\log x)^2) \ud x}
		=+\infty,
	\end{equation}
	\fi
\end{example}

% Example: h=-inf
\begin{example}
	%To show the necessity of \eqref{eqn1:defAC}, consider the following probability density function:
	Consider the following probability density function:
	\begin{equation}
		q(x)=
		\begin{cases}
			\frac{-1}{x(\log x) (\log(-\log x))^2}, & 0<x<\ue^{-\ue}, \\
			0, & \text{otherwise}.
		\end{cases}
	\end{equation}
Here, \eqref{eqn2:defAC} is satisfied but \eqref{eqn1:defAC} is violated. Note that $q(x)$ is a valid density function with no delta functionals, thus $q\in\mathcal{AC}$. Since the support of this pdf is finite, all of its $\alpha$-th moments are finite.
	However, by the change of variables $y=\log(-\log x))$, we have that
	\iffalse
	\begin{align}
		\h{q}=&\int_0^{\ue^{-\ue}}
		{\frac{-\log(-x(\log x) (\log (-\log x))^2)}{x(\log x) (\log (-\log x))^2}
		 \ud x} \nonumber\\
		=&\int_1^\infty
		{\frac{1}{\exp(-\ue^y) \ue^y y^2}
		\log\left(\exp(-\ue^y) \ue^y y^2\right)
		\exp(-\ue^y)\ue^y \ud y}\nonumber\\
		=&2\int_1^\infty{\frac{\log y}{y^2} \ud y}
		-\int_1^\infty{\frac{\ue^y-y}{y^2} \ud y}=-\infty,
	\end{align}
	where $y=\log(-\log x)$.
	\else
	\begin{equation}
		\h{q}=\int_0^{\ue^{-\ue}}
		{\frac{-\log(-x(\log x) (\log (-\log x))^2)}{x(\log x) (\log (-\log x))^2}
		 \ud x}
		 =-\infty.
	\end{equation}
	\fi
\end{example}

%%%%%
% Proofs
%%%%%
\section{Proof of the Main Result} \label{sec:Proofs}
%%%%%%%%%%%%%%%%%%%%%
% First Part: Finite Varience Entropy Existance
\subsection{Proof of \eqref{eq.dh(pd)=0}: Upper bound on the differential entropy}
Let $\bY:=\sqrt[n]{c_m}\,\bX$ where $c_m=\max(m,1)$.
The existence of the differential entropy $\h{\bY}$ is equivalent to the  existence of the differential entropy $\h{\bX}$, and when both exist we have  that \cite[Theorem 8.6.4]{Cover06},
\begin{equation} \label{eq.EntChgVar}
	\h{\bX} = \h{\bY} - \frac{1}{n}\log c_m.
\end{equation}
The benefit of defining $\bY$ is that its density function denoted by $q$, is  bounded from above by one:
\begin{equation}
	q(\by)=\frac{1}{c_m} p_{\bX}\left(\frac{\by}{\sqrt[n]{c_m}} \right) \stackrel{\rm a.e.}{\leq} 1.
\end{equation}
Hence, for any $\by\in\mathbb{R}^n$, we know that
\begin{equation} \label{eq.pxlnpxPositive}
	q(\by) \log \frac{1}{q(\by)} \geq 0.
\end{equation}
To prove the convergence of 
\begin{equation}
	\int_{\mathbb{R}^n}{q(\by) \log\frac{1}{q(\by)}\ud \by}
	=\int_{\mathbb{R}^n}{q(\by) \left|\log\frac{1}{q(\by)}\right|\ud \by},
\end{equation}
let us consider 
\begin{equation}
	\Delta_w=\int_{\lVert\by\rVert_\infty\leq w}{q(\by) \log \frac{1}{q(\by)} \ud \by}, 
\end{equation}
for $w>0$, where $\lVert\by\rVert_\infty:=\max_{i\in{1,\cdots,n}}{|y_i|}$, and $y_i$ is the $i$th element of $\by$. By recalling \eqref{eq.pxlnpxPositive}, we deduce that $\Delta_w$  is increasing in $w$.
In addition, 
\begin{equation}
	\h{q}=\h{\bY} = \lim_{w\rightarrow\infty}\Delta_w.
\end{equation}
Thus, to prove the existence of the differential entropy $\h{\bY}$, it suffices to find a constant $\kappa$ such that $\Delta_w\leq \kappa$ for all $w$.
Then, $\kappa$ would also serve as an upper bound for the entropy $\h{\bY}$.
Let
\begin{equation}
	\Theta_w = \int_{\lVert\by\rVert_\infty\leq w}{q(\by) \log \frac{q(\by)}{\phi_{n,\alpha,c_m^{\alpha/n} v}(\by)} \ud \bx},
\end{equation}
where $\phi_{n,\alpha,c_m^{\alpha/n} v}$ is defined in Definition \ref{def.PhiAlphaV}.
Except for the domain of the integral, $\Theta_w$ defines a relative entropy measure. The utility of relative entropy to bound entropy is a known technique; in particular, we link $\Theta_w$ to $\Delta_w$ by
\begin{align}
	\Theta_w =& \int_{\lVert\by\rVert_\infty\leq w}
	{q(\by)\log\frac{q(\by)\exp\big(\frac{n}{c_m^{\alpha/n}v\alpha} \left\lVert\by\right\rVert_\alpha^\alpha\big)}
	{\left[\left(\frac{n}{c_m^{\alpha/n}v\alpha}\right)^{\frac{1}{\alpha}}\frac{1}{2\Gamma\left(\frac{1}{\alpha}+1\right)}\right]^n}\ud\by} \\
	= & n\log{\left[2\left(\tfrac{c_m^{\alpha/n} v\alpha}{n}\right)^\frac{1}{\alpha}\Gamma\left(\tfrac{1}{\alpha}+1\right)\right]}
	\int_{\lVert\by\rVert_\infty\leq w}{q(\by) \ud\by} \nonumber\\
	&+\frac{n}{c_m^{\alpha/n} v\alpha}
	\int_{\lVert\by\rVert_\infty\leq w}{\lVert\by\rVert_\alpha^\alpha q(\by) \ud\by} - \Delta_w.
\end{align}
This could be rewritten as
\begin{align}
	\Delta_w
	= & \left[\frac{n}{\alpha}\log b_1+\log{c_m}\right]
	\int_{\lVert\by\rVert_\infty\leq w}{q(\by) \ud\by} \nonumber\\
	&+\frac{n}{c_m^{\alpha/n} v\alpha}
	\int_{\lVert\by\rVert_\infty\leq w}{\lVert\by\rVert_\alpha^\alpha q(\by) \ud\by}
	- \Theta_w\\
	\leq&\frac{n}{\alpha}|\log b_1| +\log{c_m}+\frac{n}{\alpha}-\Theta_w \label{eq.hn},
\end{align}	
where $b_1:={\alpha v}/{n}\left[2\Gamma\left({1}/{\alpha}+1\right)\right]^\alpha$, and the last inequality is true because
\begin{align}
	\int_{\lVert\by\rVert_\infty\leq w}{q(\by) \ud\by}
	\leq& \int_{\mathbb{R}^n}{q(\by) \ud\by}
	= 1,\\
	\int_{\lVert\by\rVert_\infty\leq w}{\lVert\by\rVert_\alpha^\alpha q(\by) \ud\by}
	\leq& \int_{\mathbb{R}^n}{\lVert\by\rVert_\alpha^\alpha q(\by) \ud\by}\\
	=&\E{\lVert\bY\rVert_\alpha^\alpha}
	=c_m^{\alpha/n} v.
\end{align}
We now find a lower bound for $\Theta_w $.
Note that for any $x\geq 0$ and $a>0$, we have
$
x\log({x}/{a}) \geq x-a.
$
Thus,
\begin{align}
	\Theta_w =&\int_{\lVert\by\rVert_\infty\leq w}
	{q(\by) \log \frac{q(\by)}{\phi_{n,\alpha,c_m^{\alpha/n} v}(\by)} \ud\by} \\
	\geq &\int_{\lVert\by\rVert_\infty\leq w}
	\left(q(\by) -\phi_{n,\alpha,c_m^{\alpha/n} v}(\by) \right)\ud\by\\
	= &\int_{\lVert\by\rVert_\infty\leq w}{q(\by) \ud\by}-\int_{\lVert\by\rVert_\infty\leq w}
	{\phi_{n,\alpha,c_m^{\alpha/n} v}(\by) \ud\by} 
	\geq -1 \label{eq.dnBnd},
\end{align}
where the validity of the last equation is coming from the fact that $\phi_{n,\alpha,c_m^{\alpha/n} v}$ and $q$ are both probability density functions, and their integrals over any interval is between $0$ and $1$. 
By combining \eqref{eq.hn} and \eqref{eq.dnBnd}, we obtain that
\begin{equation} \label{eq.hnBnd}
	\Delta_w \leq\frac{n}{\alpha}|\log{b_1}|+\log{c_m}+\frac{n}{\alpha}+1.
\end{equation}
This establishes an upper bound on $\Delta_w$ independent of $w$, which completes the proof of the existence of $\h{\bY}$, and in turn the existence of $\h{\bX}$.

%%%%%%%%%%%%%%%%%%%%%%%%%%%%
% Proof of the Second Part: Continuity of Differential Entropy
\subsection{Proof of \eqref{eqn:h-h<l1}: Continuity of the differential entropy}
The existence of $\h{\bX}$ and $\h{\bY}$ follow from \eqref{eq.dh(pd)=0}. The idea is to adapt the 
argument of  \cite[Theorem 17.3.3]{Cover06} from discrete entropy to differential entropy.  Let $\delta=\|p_X-p_Y\|_1$.
Observe that  $\delta=0$ yields $p_\bX \stackrel{\rm a.e.}{=} p_\bY$ and $\h{\bX} = \h{\bY}$.
For $\delta>0$, consider two new random vectors
\begin{equation} \label{eq.X'}
	\bX'=\sqrt[n]{m\ue}\,\bX,\qquad \bY'=\sqrt[n]{m\ue}\,\bY.
\end{equation}
Because of Lemma \ref{lmm:meX} from the appendix, $p_{\bX'},p_{\bY'}\in \left(\alpha,(m\ue)^{\alpha/n} v,{1}/{\ue}\right)$--$\mathcal{AC}^{n}$, as well as,
\begin{align}
	&\int_{\mathbb{R}^n}{\left| p_{\bX'}(\bx)-p_{\bY'}(\bx) \right| \ud\bx}
	=\delta, \label{eq.EntConv3}\\
	&\left| \h{\bX}-\h{\bY} \right| = \left|\h{\bX'}-\h{\bY'}\right|. \label{eq.EntConv4}
\end{align}
We further define a random vector $\bZ$ with density $p_{\bZ}$ as
\begin{equation} \label{eq.pZ}
	p_{\bZ}(\bx) := \frac{\left| p_{\bX'}(\bx)-p_{\bY'}(\bx) \right|}{\delta}.
\end{equation}
We shall show that $p_{\bZ}\in \left(\alpha,2(m\ue)^{\alpha/n}/\delta,2/(\ue\delta)\right)$--$\mathcal{AC}^n$. Since $p_{\bX'},p_{\bY'}$ represent absolutely continuous distributions, the same applies to $p_{\bZ}$. We further have
\begin{align}
	p_{\bZ}(\bx)
	=& \,\frac{\left| p_{\bX'}(\bx)-p_{\bY'}(\bx) \right|}{\delta} \\
	\leq & \,\frac{p_{\bX'}(\bx) + p_{\bY'}(\bx)}{\delta} \stackrel{\rm a.e.}{\leq} \frac{2}{\ue\delta}.
\end{align}
Regarding the $\alpha$ moment of $\bZ$, we can write that
\begin{align}
	\E{\left\lVert \bZ\right\rVert_\alpha^\alpha}
	=&\int_{\mathbb{R}^n}{\lVert\bx\rVert_\alpha^\alpha\frac{\left| p_{\bX'}(\bx)-p_{\bY'}(\bx) \right|}{\delta} \ud\bx} \\
	\leq & \int_{\mathbb{R}^n}{\lVert\bx\rVert_\alpha^\alpha \frac{p_{\bX'}(\bx) + p_{\bY'}(\bx)}{\delta} \ud\bx} \leq 2\frac{(m\ue)^{\frac{\alpha}{n}} v}{\delta}.
\end{align}
The latter result confirms $p_{\bZ}\in \left(\alpha,2(m\ue)^{\alpha/n}/\delta,2/(\ue\delta)\right)$--$\mathcal{AC}^n$.
Now, from \eqref{eq.dh(pd)=0}, we know that $\h{\bZ}$ exists, and
\begin{align}
	\left|\h{\bZ}\right| \leq&
	\frac{n}{\alpha}\left|\log\left(\tfrac{\alpha v}{n\delta}\right) \right|
	+\log\big({\max\left\{\tfrac{m}{\delta},1\right\}\big)} \nonumber\\
	&+n\log\left[2\Gamma\left(1+\tfrac{1}{\alpha}\right)\right]
	+\frac{n}{\alpha}+1. \label{eq:h<1}
\end{align}
This result could be simplified by applying $\delta\leq m$ as
\begin{equation} \label{eq.ThZ=0}
	\left|\h{\bZ}\right|\leq c_1  + (c_2-1) |\log\delta|,
\end{equation}
where $c_1$ and $c_2$ are given in \eqref{eqc1}, \eqref{eqc2}.
Next, we show that
\begin{equation} \label{eq.UpBnd}
	\left| \h{\bX'}-\h{\bY'} \right| \leq \delta\log\frac{1}{\delta} + \delta \h{\bZ}.
\end{equation}
This equation together with \eqref{eq.EntConv4} and \eqref{eq.ThZ=0} shall complete the proof.
In order to prove \eqref{eq.UpBnd}, we can write
\begin{align}
	&\left| \h{\bX'}-\h{\bY'} \right| \nonumber\\
	&\qquad\leq\int_{\mathbb{R}^n}
	{\left| p_{\bX'}(\bx)\log\frac{1}{p_{\bX'}(\bx)}-p_{\bY'}(\bx) \log \frac{1}{p_{\bY'}(\bx)}\right|} \ud\bx\\
	&\qquad=\int_{\mathbb{R}^n}
	{\left| p_{\bX'}(\bx) \log {p_{\bX'}(\bx)}-p_{\bY'}(\bx) \log {p_{\bY'}(\bx)}\right|\ud\bx}.
\end{align}
Lemma \ref{lmm:xlnx} in the appendix results in
\begin{align}
	&\left|p_{\bX'}(\bx) \log {p_{\bX'}(\bx)}-p_{\bY'}(\bx) \log {p_{\bY'}(\bx)}\right| \\
	&\qquad\leq \delta p_{\bZ}(\bx) \log \frac{1}{\delta p_{\bZ}(\bx)}.
\end{align}
Consequently,
\begin{align}
	\left| \h{\bX'}-\h{\bY'} \right|
	\leq& \int_{\mathbb{R}^n}{\delta p_{\bZ}(\bx) \log \frac{1}{\delta p_{\bZ}(\bx)} \ud\bx} \\
	=& \delta\log\frac{1}{\delta} + \delta\h{\bZ},
\end{align}
which completes the proof.

%%%%%%%
% Conclusion
%%%%%%%
%\section{conclusion} \label{sec:Conclusion}
%In this paper, a class of absolutely continuous probability distributions in $n$-dimensional space, $(\alpha,m,v)$--$\mathcal{AC}^n$, has studied.
%Then, it has shown that the differential entropy is uniformly convergent over this space under the total variation distance metric.

%%%%%%%
% References
%%%%%%%
\bibliographystyle{IEEEtran}
\bibliography{IEEEabrv,mybib}

%%%%%%
% Appendix
%%%%%%
\appendix
% Lemma: Y'=meX and Y'=meY
\begin{lemma} \label{lmm:meX}
	Let $\bX,\bY$ be distributed according to $p_\bX,p_\bY\in(\alpha,v,m)\text{--}\mathcal{AC}^n$, respectively.
	Let $\bX'=\sqrt[n]{m\ue}\,\bX, \bY'=\sqrt[n]{m\ue}\,\bY$ be scaled versions of $\bX$ and $\bY$ respectively.
	Then, 
	\begin{align}
		&p_{\bX'},p_{\bY'}\in\left(\alpha,(m\ue)^\frac{\alpha}{n} v,\tfrac{1}{\ue}\right)\text{--}\mathcal{AC}^n,\label{eqn:X'amvAC}\\
		&\left\|p_\bX-p_\bY\right\|_1=\left\|p_{\bX'}-p_{\bY'}\right\|_1,\label{eqn:|X'-Y'|}\\
		&\h{\bX}-\h{\bY}=\h{\bX'}-\h{\bY'}. \label{eqn:|hX'-hY'|}
	\end{align}
\end{lemma}

\begin{IEEEproof}
	For \eqref{eqn:X'amvAC}, note that
	\begin{equation} \label{eqn:pX' wrt pX}
		p_{\bX'}(\bx) = \frac{1}{m\ue} p_\bX\left(\tfrac{\bx}{\sqrt[n]{m\ue}}\right).
	\end{equation}
	Since, $p_\bX\in(\alpha,v,m)\text{--}\mathcal{AC}^n$, then $p_{\bX'}(\bx) \stackrel{\rm a.e.}{\leq} {1}/{\ue}$.
	Furthermore, we have that $\E{\left\lVert\bX'\right\rVert_\alpha^\alpha}= (m\ue)^{{\alpha}/{n}}\E{\lVert \bX \rVert_\alpha^\alpha}\leq (m\ue)^{{\alpha}/{n}} v$.
	Equation \eqref{eqn:|X'-Y'|} is immediate from the definition of the total variation distance, and \eqref{eqn:|hX'-hY'|} holds because scaling a variable by $\sqrt[n]{m\ue}$ increases its entropy by 
$({1}/{n})\log(m\ue)$.
\end{IEEEproof}

% Lemma: intxlnx
\begin{lemma}\cite[Theorem 17.3.3]{Cover06} \label{lmm:xlnx}
	For every $x,y\in[0,1/\ue]$ we have that
	\begin{equation} \label{eq.xlnxIneq}
		|x\log x - y\log y| \leq |x-y| \log \frac{1}{|x-y|}.
	\end{equation}
\end{lemma}

\iffalse
\begin{IEEEproof}
The key in this proof is that the function $f(x)=x\log x$ is both decreasing and convex in $[0,1/\ue]$. This in turn shows that $x\mapsto f(x)-f(x+a)$ is also decreasing in $x$ for any $a>0$, and hence, $f(l)-f(u)\leq f(0)-f(u-l)=-f(u-l)$, where $l=\min\{x,y\}$ and $u=\max\{x,y\}$. 

Suppose $x<y$.
Due to the fact that $f(x)=x\log x$ is decreasing  in $[0,1/e]$, we have $|f(x)-f(y)|=f(x)-f(y)$. Since $f(x)$ is convex, $x\mapsto f(x)-f(x+a)$ is decreasing in $x$ for any $a>0$, and hence $f(x)-f(y)\leq f(0)-f(y-x)=-f(y-x)$.
\end{IEEEproof}
\fi

\end{document}